\documentclass[10pt]{iopart}

\expandafter\let\csname equation*\endcsname=\relax
\expandafter\let\csname endequation*\endcsname=\relax
\usepackage{amsmath}
\usepackage[english]{babel}
\usepackage{iopams}  
\usepackage{array} 
\usepackage{graphicx} 
\usepackage{amstext}
\usepackage{amsbsy}
\usepackage{txfonts}
\usepackage{subfigure}
\usepackage{multirow}
\usepackage{hyperref}
\usepackage{microtype}
\usepackage{caption}
\usepackage{booktabs} 
\usepackage{cite}
\newcolumntype{C}[1]{>{\centering\arraybackslash}p{#1}}
\begin{document}

\title{Plasma Shape Control via Zero-shot Generative Reinforcement Learning}

\author{Niannian Wu$^{1}$, Rongpeng Li$^{1,*}$, Zongyu Yang$^{2}$, Yong Xiao$^{1}$, Ning Wei$^{3}$, Yihang Chen$^{2}$, Bo Li$^{2}$, Zhifeng Zhao$^{3}$, and Wulyu Zhong$^{2,*}$}
\address{1. Zhejiang University, Hangzhou 310058, China}
\address{2. Southwestern Institute of Physics, Chengdu 610043, China}
\address{3. Zhejiang Lab, Hangzhou 311500, China}
\ead{lirongpeng@zju.edu.cn, zhongwl@swip.ac.cn}
\vspace{10pt}

\begin{abstract}
Traditional PID controllers have limited adaptability for plasma shape control, and task-specific reinforcement learning (RL) methods suffer from limited generalization and the need for repetitive retraining. To overcome these challenges,  
this paper proposes a novel framework for developing a versatile, zero-shot control policy from a large-scale offline dataset of historical PID-controlled discharges. Our approach synergistically combines Generative Adversarial Imitation Learning (GAIL) with Hilbert space representation learning to achieve dual objectives: mimicking the stable operational style of the PID data and constructing a geometrically structured latent space for efficient, goal-directed control. The resulting foundation policy can be deployed for diverse trajectory tracking tasks in a zero-shot manner without any task-specific fine-tuning. Evaluations on the HL-3 tokamak simulator demonstrate that the policy excels at precisely and stably tracking reference trajectories for key shape parameters across a range of plasma scenarios. This work presents a viable pathway toward developing highly flexible and data-efficient intelligent control systems for future fusion reactors.

\vspace{2pc}
\end{abstract}


%
%
\ioptwocol

\section{Introduction}
The development of sustainable clean energy represents one of the most critical scientific challenges of our time, and magnetic confinement fusion, particularly in tokamak devices, stands as a primary technological pathway toward this goal. A core prerequisite for achieving the stable, high-performance plasma discharges necessary for net energy gain is the precise and robust control of the plasma's shape and current. The ability to dynamically regulate key geometric parameters is essential for maintaining plasma stability \cite{nelson2023vertical, inoue2023adaptive}, optimizing performance \cite{hofmann1990plasma,reimerdes2000effect}, and preventing disruptive events \cite{samm1999marfe,holcomb2009optimizing} that can damage the reactor wall. Traditionally, tokamak plasma shape control has been dominated by Proportional-Integral-Derivative (PID) controllers \cite{de2019plasma, mitrishkin2020two}. 
While this method has proven reliable for maintaining basic configurations, its inherent limitations, such as laborious manual parameter tuning and difficulty in coping with abrupt perturbations \cite{blum2019automating, de2019plasma}, are significant. 
In this regard, deep reinforcement learning (DRL) has emerged as a powerful alternative, and achieved notable breakthroughs, including precise control of complex plasma shapes \cite{degrave2022magnetic, wu2025high}, real-time tracking of key kinetic parameters \cite{seo2021feedforward,char2023offline}, and active suppression of performance-degrading instabilities \cite{seo_avoiding_2024}. 

While these advances showcase the formidable capability of RL in addressing specific control problems, two critical deficiencies in the current methodology hinder the development of a qualified zero-shot RL-based plasma shape controller. Here, similar to the counterparts in natural language processing and computer vision \cite{radford2018improving, ge2023policy}, a zero-shot RL agent shall have the capability of tracking new, unforeseen reference trajectories \cite{reed2022a, han2024lifelike}, which is essential for unpredictable disruption mitigation events \cite{seo_avoiding_2024}. However, the conventional RL agent \cite{seo_avoiding_2024,degrave2022magnetic,wu2025high} is commonly single-task oriented or determined by a carefully engineered reward function. For instance, in \cite{seo_avoiding_2024}, different safety thresholds for tearing mode avoidance necessitate the training of distinct high-level RL controllers. Similarly, in the plasma shape control studies by Degrave \textit{et al.} \cite{degrave2022magnetic} and Wu \textit{et al.} \cite{wu2025high}, distinctive target plasma shapes require the training of separate policies. While more advanced approaches can train a single constraint-conditioned policy to adapt to varying operational limits \cite{wang2025active}, this flexibility is bound to a hardcoded final target, such as ramping down the plasma current to below $2$ MA. Consequently, changing this pre-defined target to a different final current threshold would necessitate retraining the entire policy, which does not grant the ability to pursue new targets in a zero-shot manner. This suggests that fundamental changes in control objectives or task scenarios will inevitably trigger the redesign of the reward function and retraining of a new policy, as well as the need for cumbersome data cleansing, resulting in significant costs for data preparation and computational resources. Moreover, the need for new control targets is persistent and multifaceted, as most current and future devices, including ITER, retain a crucial role in scientific exploration, meaning control objectives are inevitably adjusted during repetitive discharges. Compounding this, to cope with possible disruptions during a discharge, the required shape adjustments could be rather unpredictable. In this case,   
a single, fixed shape control-oriented RL agent becomes less qualified than well-calibrated PID controllers \cite{seo_avoiding_2024}.
On the other hand, due to the high cost and difficulty of training RL algorithms through direct interaction with physical devices, current work in RL-based control predominantly follows a model-based paradigm. 
This typically involves either relying on physics-based simulators \cite{degrave2022magnetic} or constructing surrogate models from vast historical datasets \cite{seo_avoiding_2024,wu2025high,seo2021feedforward}. 
The adopted simulated environment inevitably contains certain 
gap with the practical, physical device, 
and yield learned policies that are Out-Of-Distribution (OOD) \cite{agarwal2020optimistic} or potentially destructive when deployed in the real world. For example, in experiments on the KSTAR tokamak, Seo \textit{et al.} \cite{seo2021feedforward} find that their surrogate model's inability to predict fast MHD instabilities leads the trained policy to drive the plasma into a physically unstable state, ultimately triggering a large-scale tearing mode.

To build a zero-shot RL agent, Generative Adversarial Imitation Learning (GAIL) \cite{ho2016generative}, which learns a diverse set of fundamental skills from large, unlabeled offline datasets, offers an effective solution. In the field of physics-based character animation, GAIL algorithms, such as AMP \cite{peng2021amp}, ASE \cite{peng2022ase}, and CALM \cite{tessler2023calm}, first leverage imitation learning to extract a variety of modular behavioral patterns (e.g., walking and jumping) from expert data. Through adversarially confining the learned policy to a data-validated, safe operational manifold, GAIL inherently mitigates the risk of exploiting simulator inaccuracies to maximize its reward, thus significantly reducing ineffective or hazardous OOD actions. On this basis, a low-dimensional latent space encoding complex task objectives serves as the anchor for further mobilizing appropriate modular patterns. Consequently, despite its effectiveness in the behavior-limited animation field, the separate learning nature of individual modular behaviors required in GAIL implies additional training for a new behavior and limits the capability of zero-shot generalization. This deficiency is also compounded by the complexity of plasma shape control, a far more diverse scenario with a wide range of plasma configurations (e.g., reference elongations and triangularities) \cite{hofmann1990plasma, reimerdes2000effect} and current profiles \cite{ou2009controllability, ding2016scenario}.

In this paper, we propose a novel framework for zero-shot plasma shape control in HL-3 tokamak.
Particularly, our approach aims to learn safe operating policies as in GAIL-based frameworks. Nevertheless, to collectively learn multiple configurations, distinct from GAIL, we develop a geometrically structured Hilbert latent space \cite{park2024foundation}. In this space, the geometric distance between any two points directly corresponds to the optimal reachability between different states (i.e., plasma currents and shapes) in the physical environment, enabling synergistic learning. 
In other words, instead of repetitively learning multiple modular policies, our approach effectively learns an effective representation in the Hilbert latent space as a navigational map to guide the agent toward the target configuration, making the plasma shape control problem for new configurations analytically solvable. The simultaneous achievement of dual objectives (i.e., data mimicking and Hilbert policies) allows for the direct utilization of a large-scale, unlabeled dataset containing diverse PID-controlled discharges, rather than laboriously pre-classifying the data according to operational modes. Finally, the unanimously pre-trained RL policy can naturally support new shape control tasks over a wide range of operational scenarios, including diverse plasma shapes and multiple plasma currents, in a zero-shot manner. Extensive simulation results validate the effectiveness of the proposed method.

This paper is organized as follows. Section \ref{sec:methodology} details our methodology, covering the dataset preparation, model architecture, and the overall training strategy. In Section \ref{sec:results}, we present the experimental results, beginning with an evaluation of the controller's tracking control performance during the flat-top phase of different plasma current and shape scenarios, followed by an analysis of the learned latent space. Finally, Section \ref{sec:discussion} concludes the paper with a summary and discussion.

\section{Methods}
\label{sec:methodology}
\subsection{Dataset Construction and Preprocessing}
The HL-3 Tokamak, designed and constructed by the Southwestern Institute of Physics (SWIP) of the China National Nuclear Corporation (CNNC), is an advanced tokamak facility. It features a major radius $R = 1.78$ m, a minor radius $a = 0.65$ m, and an aspect ratio of $2.8$. The device is designed for a plasma current ($I_p$) in the range of $2.5-3$ MA and a toroidal magnetic field ($B_t$) of $2.2-3$ T \cite{zhong2024china}. The poloidal field configuration of HL-3 is intricately managed by a Central Solenoid, complemented by an ensemble of eight pairs of poloidal field coils strategically distributed as inner coils (PF1-PF4), top-bottom coils (PF5-PF6), and outer coils (PF7-PF8), with the latter trio playing a pivotal role in establishing the divertor configuration, hence their nomenclature as divertor coils \cite{song2021plasma, song2019first, xue2019integrated}.

\begin{table*}[!tb]
\centering
\captionsetup{justification=centering}
\caption{\label{tab:channels}A summary of the variables for model training.}
\begin{tabular*}{0.9\textwidth}{@{} c @{\extracolsep{\fill}} c @{\extracolsep{\fill}} c @{\extracolsep{\fill}} c @{} c @{\extracolsep{\fill}}c @{}}
\br
Variable&Unit&Channels&Description&Source\\
\mr
$I_p$ & kA & $1$ & Plasma current & Magnetics \\
$a$ & cm & $1$ & Minor radius & EFITNN \\
$\kappa$ & dimensionless & $1$ & Elongation & EFITNN \\
$\delta_u$ & dimensionless & $1$ & Upper triangularity & EFITNN \\
$\delta_l$ & dimensionless & $1$ & Lower triangularity & EFITNN \\
$R$ & cm & $1$ & Radial position of plasma geometric center & EFITNN \\
$Z$ & cm & $1$ & Vertical position of plasma geometric center & EFITNN \\
$I_c$ & A & $18$ & Coil current & Magnetics \\
$U$ & V & $17$ & Coil voltage & Magnetics \\
$TF$ & T & $1$ & Toroidal magnetic field & Magnetics \\
$V_{\text{loop}}$ & V & $1$ & Loop voltage & Magnetics \\
\br
\end{tabular*}
\end{table*}

\begin{figure*}[!ht]
    \centering
    \includegraphics[width=0.9\textwidth]{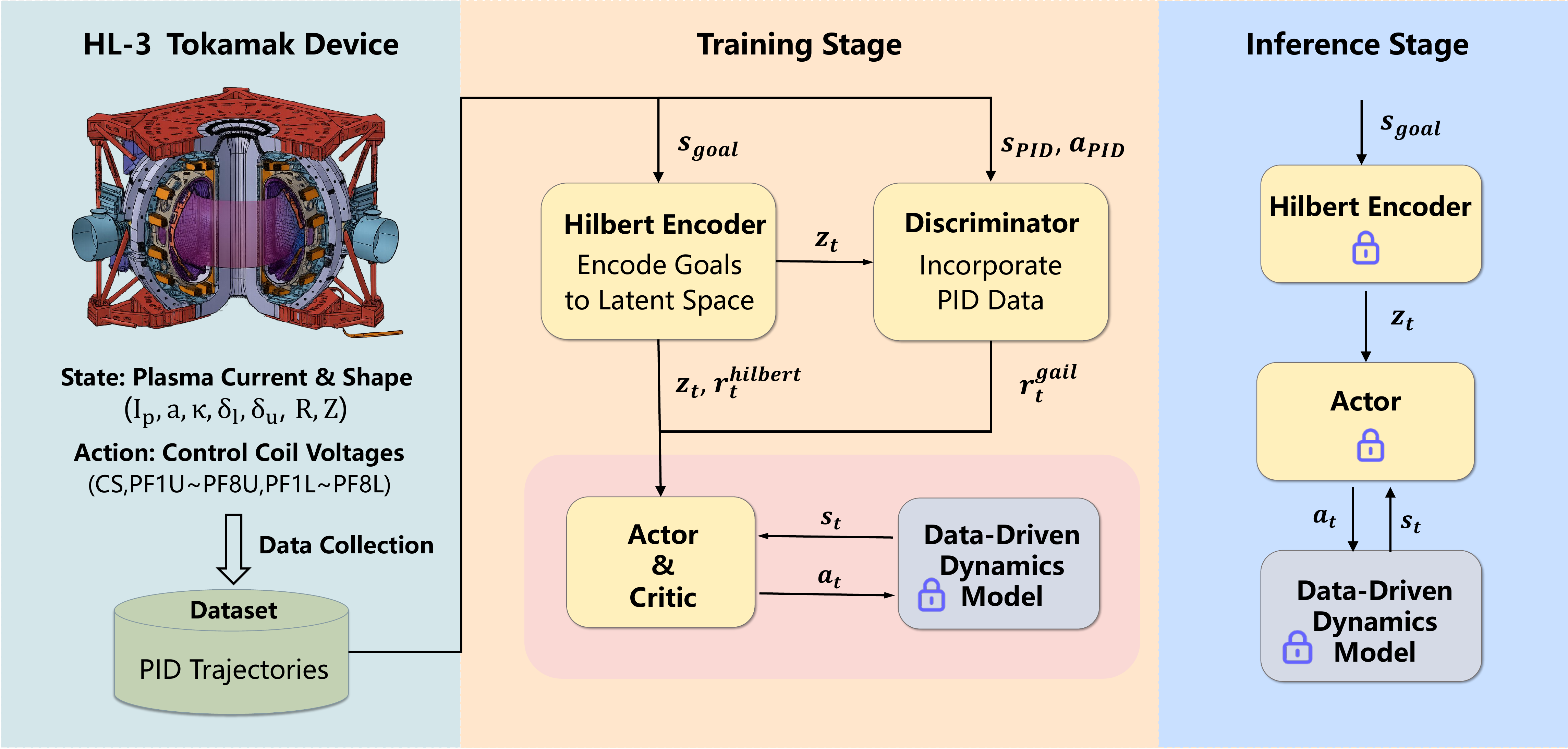}
    \caption{\textbf{Overview of the proposed generative reinforcement learning framework for plasma current and shape control on the HL-3 Tokamak System.}
    The framework consists of a data collection module, a training stage, and an inference stage. In the training phase, a Hilbert Encoder maps goals to a structured latent space and generates intrinsic directional rewards. A discriminator incorporates PID data to provide imitation rewards, and an RL Policy is trained to generate control actions based on these intrinsic and imitation rewards. This training is conducted by having the agent interact with a pretrained data-driven dynamics model, which serves as a simulation environment. During the inference stage, the trained Hilbert encoder and RL agent are deployed to achieve zero-shot control over diverse reference plasma configurations.}
    \label{fig:framework}
\end{figure*}
For this study, we use a dataset of $1,115$ valid discharges, shots \#5000--\#11908, produced under PID control on the HL-3 tokamak during the 2023--2025 experimental campaigns. We select shots where the plasma current is continuously maintained above $100$ kA. This current threshold has been established to mitigate computational errors in plasma shape reconstruction that can arise from eddy current effects in passive conductors, particularly during the low-current phase (e.g., the initial ramp-up). Notably, this encompasses a diverse range of plasma scenarios and rich plasma configurations, including limiter, divertor, and advanced divertor discharges. However, it excludes failed discharges and those terminating in major disruptions. Meanwhile, the dataset comprises $441$ shots with a plasma current flat-top of $300$ kA, $473$ shots at $500$ kA, and $201$ shots at $600$ kA. Each shot contains $44$ data channels, comprising $17$ control coil voltages, $18$ control coil currents, $1$ poloidal field coil current, $1$ loop voltage, $1$ plasma current, and $6$ plasma shape parameters. Detailed channel information is provided in Table~\ref{tab:channels}. To eliminate the influence of different physical units across variables, we apply min-max normalization to each channel:
\begin{equation}
x_{\text{norm}} = \frac{x - x_{\min}}{x_{\max} - x_{\min}}
\label{eq:normalization}
\end{equation}
where $x_{\min}$ and $x_{\max}$ are the minimum and maximum values, respectively, statistically determined for each channel across all $1,115$ shots.

\begin{table}[t]
\centering
\footnotesize
\caption{\centering The state and action space definition for RL. \label{tab:state and action space}}
\centering{}
\begin{tabular}{@{}C{1.8cm}C{5.5cm}}
\br
RL variable&Description\\
\mr
\multirow{1}{*}{State}&$\left(I_{p}, a, \kappa, \delta_{l}, \delta_{u},  R, Z\right)$\\
\multirow{1}{*}{Action}&{$\left( {CS}, PF1U \sim PF8U, PF1L\sim PF8L\right)$}\\
\br

\end{tabular}
\end{table}
\subsection{Generative RL Model}
Contingent on a ready simulator of HL-3 tokamak, we aim to learn a zero-shot RL controller, which possesses the control capabilities of the $17$ control coil voltages for regulating plasma current ($I_p$) and six key shape parameters ($a$, $\kappa$, $\delta_l$, $\delta_u$, $R$ and $Z$) towards any reference configurations, while adhering to the operational characteristics of the HL-3 tokamak.
Notably, consistent with our previous work \cite{wu2025high}, the RL state, denoted as $s \in \mathbb{R}^7 $, is a $7$-dimensional vector composed of the plasma current ($I_p$) and six key geometric parameters: the minor radius ($a$), elongation ($\kappa$), upper triangularity ($\delta_u$), lower triangularity ($\delta_l$), and the radial ($R$) and vertical ($Z$) coordinates of the plasma's geometric center. Meanwhile, the action $ a \in \mathbb{R}^{17}$, is a $17$-dimensional vector representing the voltages applied to the $17$ active control coils, as summarized in Table~\ref{tab:state and action space}. Different from \cite{wu2025high}, wherein the reward takes direct deviations between the current plasma state $s$ and a particular reference state $s_\text{goal}$, we aim to bypass the underlying reward engineering and policy re-training and develop a zero-shot RL agent to track multiple new target configurations or currents capably.

The overall architecture of our proposed framework is illustrated in Figure~\ref{fig:framework}. Apart from an actor $\pi_{\psi}$ and a critic $V_{\omega}$, the framework also encompasses a Hilbert encoder $E_{\theta}$ and an adversarial discriminator $D_{\phi}$. The training of this framework is orchestrated around two key datasets: a static, offline dataset of expert trajectories, denoted as $\mathcal{D}_{\rm PID}$, containing stable historical data from the PID controller; and an on-policy dataset, $\mathcal{D}_{\rm RL}$, populated with experience collected as the policy interacts with the dynamics model. These components and data sources work in synergy. 
Notably, the Hilbert encoder $E_{\theta}$ maps target trajectories $s_\text{goal}$ into a geometrically structured latent space representation $z_\text{goal}$. Within this learned latent space, a directional reward $r_\text{hilbert}$ can be more easily derived from the distance between the current latent state and the target latent representation (i.e., $z$ and $z_\text{goal}$), providing continuous guidance towards any desired configuration. Meanwhile, the adversarial discriminator $D_{\phi}$ constrains the RL policy to a data-validated, safe operational manifold by providing an imitation reward $r_\text{gail}$. This reward stems from an adversarial process in which the discriminator is trained to differentiate between trajectories generated by the actor's behavior $\mathcal{D}_{\rm RL}$ and the PID discharges $\mathcal{D}_{\rm PID}$, 
while the actor is simultaneously rewarded for generating trajectories that are indistinguishable from the PID data $\mathcal{D}_{\rm PID}$.

\subsubsection{Hilbert Encoder}
\begin{figure}[!ht]
    \centering
    \includegraphics[width=\columnwidth]{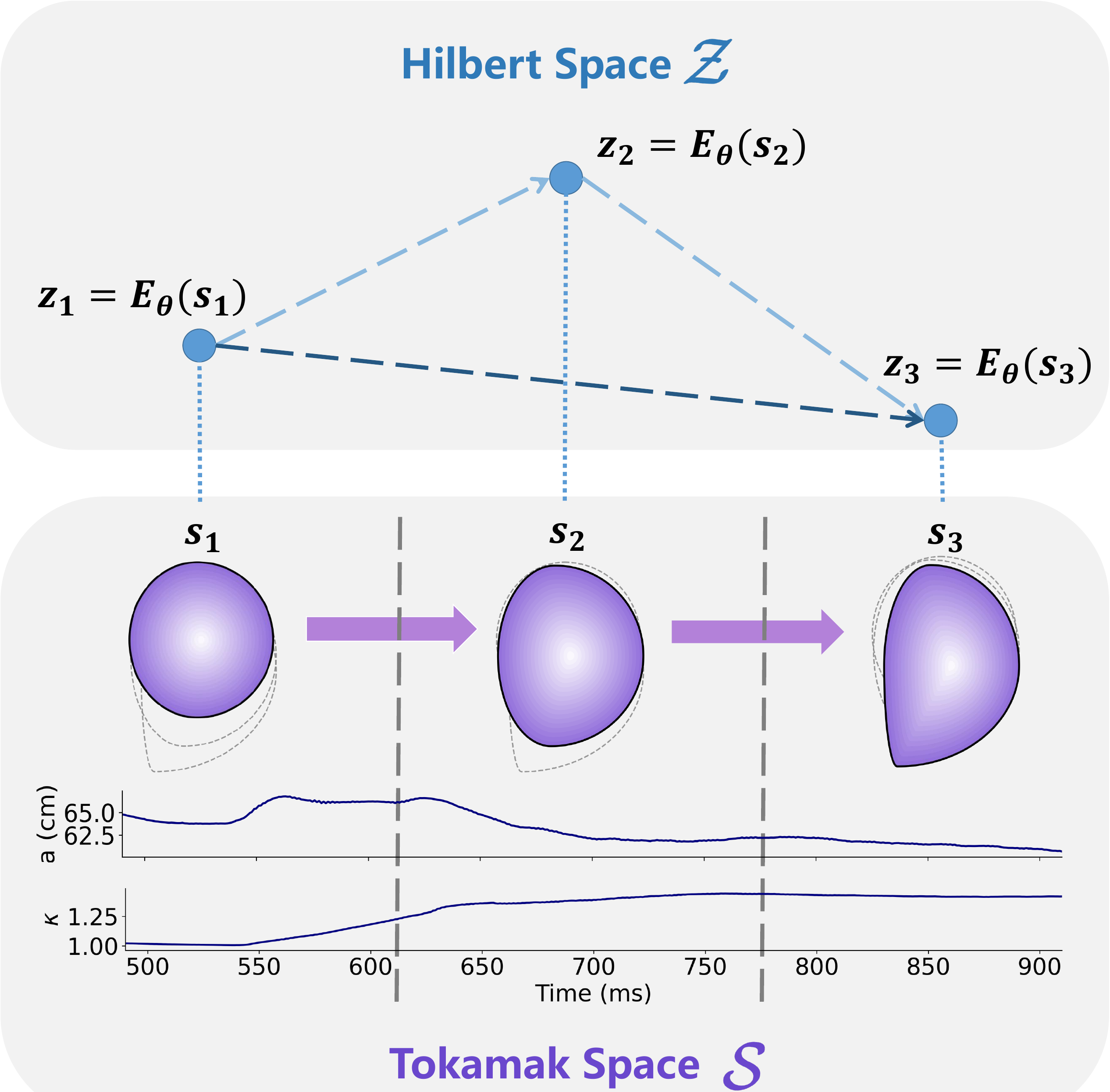}
    \caption{\textbf{Correspondence between physical plasma states and the Hilbert latent space.}
    }
    \label{fig:Hilbert Space}
\end{figure}
The primary function of the encoder $E_{\theta}$ is to map a target plasma state $s_\text{goal}$ to a latent space representation $z = E_{\theta}(s_\text{goal}) \in \mathbb{R}^d$. This encoder is implemented as a multi-layer perceptron (MLP) with ReLU as the non-linear activation function. Unlike a standard encoder, it is designed to learn a Hilbert space with a specific geometric structure. In this Hilbert space, the Euclidean distance between any two points, $z_1 = E_{\theta}(s_1)$ and $z_2 = E_{\theta}(s_2)$, is intended to approximate the optimal time-to-reach from state $s_1$ to state $s_2$ within the tokamak environment. This core concept is illustrated in Figure~\ref{fig:Hilbert Space}. This property provides the latent space with navigability, laying a foundation for the policy network to plan effective paths toward a goal. 
The training of the encoder is driven by a composite objective, with its update gradients originating from three sources: the Hilbert representation loss $\mathcal{L}_{\rm Hilbert}$, the latent space regularization loss $\mathcal{L}_{\rm reg}$, and the policy gradients loss backpropagated form the actor and critic $\mathcal{L}_{\rm policy}$, and can be expressed as:
\begin{equation}
    \mathcal{L}_{\rm Enc}(\theta) = \mathcal{L}_{\rm Hilbert} + 
    \underbrace{\mathcal{L}_{\rm consistency} + \mathcal{L}_{\rm diversity}}_{\mathcal{L}_{\rm reg}} + 
    \underbrace{\mathcal{L}_{\rm actor} + \mathcal{L}_{\rm critic}}_{\mathcal{L}_{\rm policy}}.
    \label{eq:enc_loss}
\end{equation}

Primarily, we design the Hilbert representation loss $\mathcal{L}_{\rm Hilbert}$ to impose a strict geometric constraint on the latent space, thus ensuring the Euclidean distance between any two points (e.g., $z_1 = E_{\theta}(s_1)$ and $z_2 = E_{\theta}(s_2)$) in the latent space directly corresponds to the time-optimal evolution from any two states $s_1$ to $s_2$ within the physical tokamak environment. Specifically, we model the problem as learning a goal-conditioned value function, and compute it as the negative Euclidean distance in the latent space $V(s, s_\text{goal})=-\left\|E_\theta(s)-E_\theta(s_\text{goal})\right\|$. Subsequently, to ensure the value function satisfies its Bellman equation, which describes the value relationship for an adjacent state pair $(s,s')$, we formulate a Bellman error. The Bellman error is the difference between the target value and the current value. Assuming a constant reward of $-1$ for each time step, this error is formulated as:
\begin{equation}
    \underbrace{\left(-1-\gamma_H\left\|E_{\bar{\theta}}\left(s^{\prime}\right)-E_{\bar{\theta}}\left(s_{\text {goal}}\right)\right\|\right)}_{\text{TD target}} - 
    \underbrace{\left(-\left\|E_\theta(s)-E_\theta\left(s_{\text {goal }}\right)\right\|\right)}_{\text{Current Value}}.
    \label{eq:Bellman Error}
\end{equation}
The loss $\mathcal{L}_{\rm Hilbert}$ is then constructed to minimize this Bellman error using an asymmetric quadratic loss via expectile regression \cite{newey1987asymmetric}: 
\begin{align}
\mathcal{L}_{\mathrm{Hilbert}} &= \mathbb{E}_{(s, s_{\text{goal}}, s') \sim \mathcal{D}_{\mathrm{PID}}} \Biggl[ L_2^\tau \biggl( -1 - \gamma_H \| E_{\bar{\theta}}(s') - E_{\bar{\theta}}(s_{\text{goal}}) \| \nonumber \\
&\quad + \| E_\theta(s) - E_\theta(s_{\text{goal}}) \| \biggr) \Biggr],
\end{align}
where the expectile loss $L_2^\tau$ with the hyperparameter $\tau = 0.9$ assigns greater weight to positive errors than to negative errors. This encourages the value function to be updated upwards, striking a balance between preventing significant underestimation of true value and avoiding excessive optimism in predictions. $E_{\bar{\theta}}$ is a slowly updated target encoder used to stabilize training, and $\gamma_H$ denotes a discounting constant. The incorporation of the Hilbert representation loss in \eqref{eq:enc_loss} makes our framework fundamentally different from traditional GAIL-based methods \cite{peng2021amp,peng2022ase,tessler2023calm}. 

To further improve the structural quality of the latent space, we introduce a regularization term $\mathcal{L}_{\rm reg}$, which consists of a diversity loss $\mathcal{L}_{\rm diversity}$ and a consistency loss $\mathcal{L}_{\rm consistency}$ \cite{chen2020simple}. In particular, the diversity loss $\mathcal{L}_{\rm diversity}$ promotes a uniform distribution of representations, preventing collapse to a few points, by maximizing the distance between randomly sampled pairs of points in the latent space: 
\begin{equation}
    \mathcal{L}_{\rm diversity} = \log \mathbb{E}_{z_1=E_{\theta}(s_1), z_2 =E_{\theta}(s_2)}\left[e^{-t\left\|z_1-z_2\right\|_2^2}\right],
\end{equation}
where the temperature hyperparameter $t$ is set to $2$. Conversely, the consistency loss $\mathcal{L}_{\rm consistency}$ preserves the local topology of the state space by minimizing the latent space distance between pairs of similar samples from the original space, such as consecutive time steps from the same discharge:
\begin{equation}
    \mathcal{L}_{\rm consistency} = \mathbb{E}_{(s_1, s_2) \sim \mathcal{P}_{\rm sim}}\left[\left\|E_{\theta}(s_1)-E_{\theta}(s_2)\right\|_2^2\right],
\end{equation}
where $\mathcal{P}_{\rm sim}$ is the distribution of similar sample pairs from discharges with the same plasma current flat-top and similar configurations.

In addition, to ensure this latent space is effective for the control task, the encoder $E_{\theta}$ is also updated end-to-end via policy gradients loss $\mathcal{L}_{\rm policy}$. The design of $\mathcal{L}_{\rm policy}$ will be elaborated later in Section \ref{sec:actor_critic}.

The latent variable $z$ produced by the encoder serves as a key goal-conditional input for the subsequent actor, critic, and adversarial discriminator networks. For example, for the actor $\pi_{\psi}$ and a critic $V_{\omega}$, the conditional input $z$ enables dynamic policy adjustment and value evaluation based on varying objectives, facilitating the development of a universal policy that can generalize to diverse target trajectories. Furthermore, for the discriminator $D_{\phi}$, the input triplet $(s,a,z)$ allows learning a conditional probability distribution $p(s, a | z)$. This design ensures that the adversarial imitation learning process is goal-directed, making the policy not only emulate the stable style of the PID discharges but also exhibit behavioral patterns aligned with specific objectives.

\subsubsection{Discriminator}
Given the triplet of the current state, the executed action, and the goal encoding $(s, a, z)$, the discriminator $D_{\phi}$ operates as  a binary classifier to distinguish the agent's behavior from the dynamics of a stable discharge dataset $\mathcal{D}_{\rm PID}$. It yields a scalar value between $0$ and $1$, indicating the probability that the triplet originates from the PID-controlled dataset $\mathcal{D}_{\rm PID}$, thus incentivizes the actor network to replicate the safe control style.
We employ the mean squared error (MSE) for its loss function $\mathcal{L}_\text{Disc}$, supplemented with a gradient penalty and weight regularization $\mathcal{L}_{\rm reg}(D_{\phi})$ to stabilize the training process. The loss function is defined as:
\begin{eqnarray}
    \mathcal{L}_\text{Disc}(\phi) &=& \mathbb{E}_{(s,a,z) \sim \mathcal{D}_{\rm RL}} \left[(D_{\phi}(s,a,z-0))^2\right] \nonumber\\
    && + \mathbb{E}_{(s,a,z) \sim \mathcal{D}_{\rm PID}}\left[(D_{\phi}(s,a,z)-1)^2\right] + \mathcal{L}_{\rm reg}(D_{\phi}). 
\end{eqnarray}
Within the term $\mathcal{L}_{\rm reg}(D_{\phi})$, we adopt two distinct regularization techniques to stabilize the training process and prevent overfitting. The first is a gradient penalty, which penalizes the squared $L2$ norm of the discriminator's output gradient with respect to its inputs on PID samples. This mechanism is crucial for maintaining stable training dynamics by preventing the discriminator's gradients from exploding or vanishing \cite{gulrajani2017improved}. The second technique is weight decay, implemented by imposing a penalty calculated as the Frobenius norm of all parameters in the discriminator network $\phi$, thus preventing overfitting and enhancing generality.

\begin{table*}[ht]
\centering
\caption{Model hyperparameters.}
\label{tab:model_hyperparameters_full}
\begin{tabular}{lcccc}
\toprule
\textbf{Hyperparameter} & \textbf{Actor Model} & \textbf{Critic Model} & \textbf{Discriminator Model} & \textbf{Encoder Model} \\
\midrule
Input dimension & $23$ & $23$ & $40$ & $7$ \\
Output dimension & $17$ & $1$ & $1$ & $16$ \\
Hidden layers & $3$ & $3$ & $3$ & $3$ \\
Hidden layer width & $128$ & $128$ & $128$ & $128$ \\
Input MLP width & $64$ & $64$ & $64$ & -- \\
Learning Rate & $3 \times 10^{-4}$ & $3 \times 10^{-4}$ & $1 \times 10^{-4}$ & $3 \times 10^{-4}$ \\
Output activation & Tanh & -- & Sigmoid & L2 Normalization \\
\midrule
Dropout rate & \multicolumn{4}{c}{$0.1$} \\
Activation function & \multicolumn{4}{c}{ReLU} \\
Optimizer & \multicolumn{4}{c}{Adam} \\
Learning rate scheduler & \multicolumn{4}{c}{ExponentialLR ($\gamma=0.99$)} \\
\bottomrule
\end{tabular}
\end{table*}
\subsubsection{Actor and Critic}
\label{sec:actor_critic}
As the decision-making core of our framework, the RL agent is implemented on top of the Proximal Policy Optimization (PPO) \cite{schulman2017proximal} algorithm, a state-of-the-art actor-critic method known for its stability and sample efficiency. 
Specifically, the actor network $\pi_{\psi}$ functions conditioned on the plasma state $s$ and the target shape representation $z$ generated by the encoder, outputting a Gaussian policy distribution $\pi_{\psi}(a|s,z)$ over the $17$ control coil voltages. Complementing this, the critic network $V_{\omega}$ conducts long-term value assessment. Using the same inputs $(s,z)$, it predicts the future cumulative return or state-value function $V_{\omega}(s,z)$, achievable by adhering to the current policy $\pi_{\psi}$. 

Similar to other RL approaches, after executing each action, the agents receive a reward. Here, we adopt a hybrid reward function, composed of an imitation reward $r_t^{\rm gail}$ from the discriminator and a directional reward $r_t^{\rm hilbert}$ derived from the Hilbert space:
\begin{equation}
    r_t = w_g r_t^{\rm gail} + w_h r_t^{\rm hilbert},
\end{equation}
where the terms $w_g$ and $w_h$ are the corresponding weighting coefficients. To enhance the model's sensitivity to the goal-directed signal, we set these hyperparameters to $w_g=1$ and $w_h=2$, placing a greater emphasis on efficient navigation toward the target. Traditionally, the generative adversarial imitation reward function $r_t^{\rm gail}$ is set to $-\log(1-D_{\phi}(s,a,z))$ \cite{ho2016generative}. Nevertheless, such a setting leads to policy saturation when the discriminator's output approaches $1$ and provides an insufficient penalty when the output is near $0$. Therefore, we adopt a symmetric logit function as the reward instead:
\begin{equation}
    r_t^{\rm gail} = \log(D_{\phi}(s,a,z)) - \log(1-D_{\phi}(s,a,z)).
\end{equation}
By providing substantial positive rewards for high-quality actions while imposing strong negative penalties for low-quality actions, this symmetric reward yields a more balanced and robust gradient signal for policy optimization.
On the other hand, the directional reward $r_t^{\rm hilbert}$ is derived by projecting the agent's actual displacement in the Hilbert latent space onto the direction of the goal:
\begin{equation}
    r_t^{\rm hilbert} = \frac{(E_{\theta}(s_{t+1}) - E_{\theta}(s_t)) \cdot (E_{\theta}(s_\text{goal}) - E_{\theta}(s_t))}{\|E_{\theta}(s_\text{goal}) - E_{\theta}(s_t)\|_2}.
\end{equation}
This reward component incentivizes the controller to find the most efficient path to drive the plasma from the next state $s_{t+1}$ to the goal-conditioned target state $s_\text{goal}$. However, due to its symmetric logit structure, the generative adversarial imitation reward $r_t^{\rm gail}$ is theoretically unbounded, approaching positive or negative infinity as the discriminator's output approaches its limits. This wide reward range can lead to high variance that destabilizes the training process. To solve this, we apply online reward scaling \cite{engstrom2020implementation} to the total hybrid reward $r_t$ to ensure stable learning process.

Based on the immediate hybrid reward $r_t$, we compute the advantage function $A_t$ and cumulatively discounted returns $R_t$.  
Specifically, we use the Generalized Advantage Estimator (GAE) \cite{schulman2016high}, which provides a stable estimate of the advantage:
\begin{equation}
A_t = \sum_{l=0}^{T-t-1} (\gamma\lambda)^l \delta_{t+l},
\end{equation}
where $\delta_t$ is the temporal difference (TD) error given by $\delta_{t}=r_t+\gamma V_{\omega}(s_{t+1}, z_{t+1})-V_{\omega}(s_{t}, z_{t})$, and $\lambda$ is the GAE hyperparameter set to $0.95$. 
The discounted return $R_t$ represents the cumulative sum of all future rewards from the current time step $t$ onward, with future rewards being valued less than immediate ones. It is formally expressed as:
\begin{equation}
R_t = \sum_{k=0}^{T-t} \gamma^k r_{t+k},
\end{equation}
where $\gamma$ is the discount factor set to $0.98$.
\begin{figure*}[!ht]
    \centering
    \includegraphics[width=0.95\textwidth]{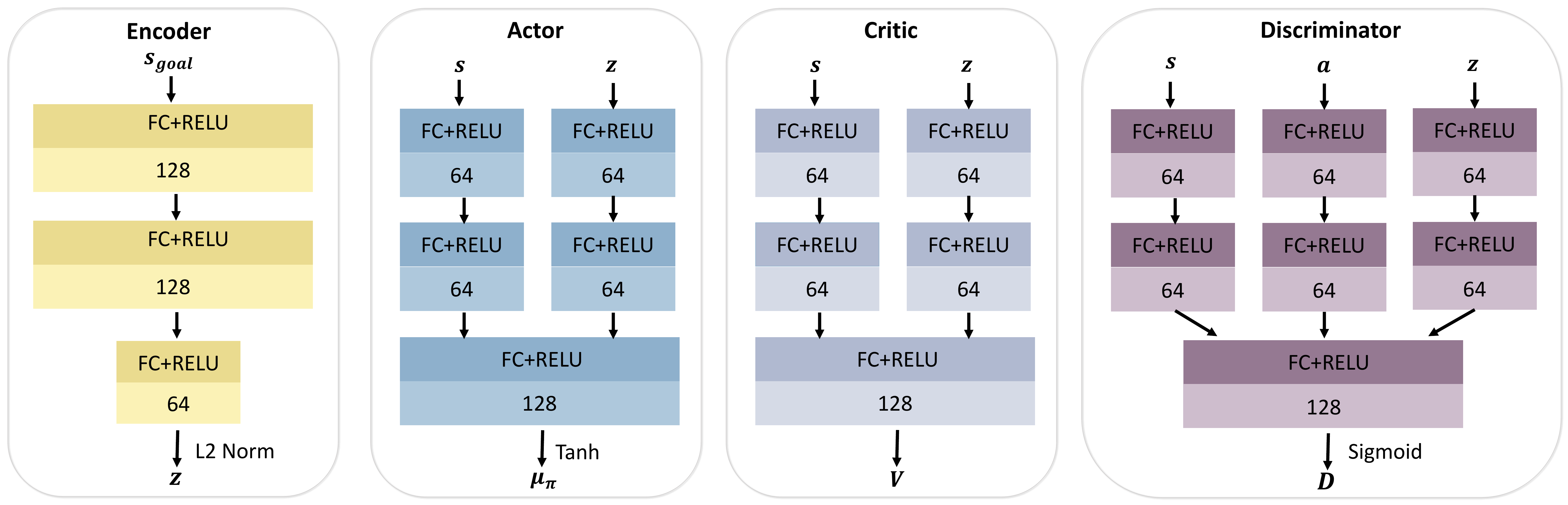}
    \caption{\textbf{Network architectures for the Encoder, Actor, Critic, and Discriminator modules.} 
    }
    \label{fig:model_network}
\end{figure*}
\begin{figure*}[!ht]
    \centering
    \includegraphics[width=0.9\textwidth]{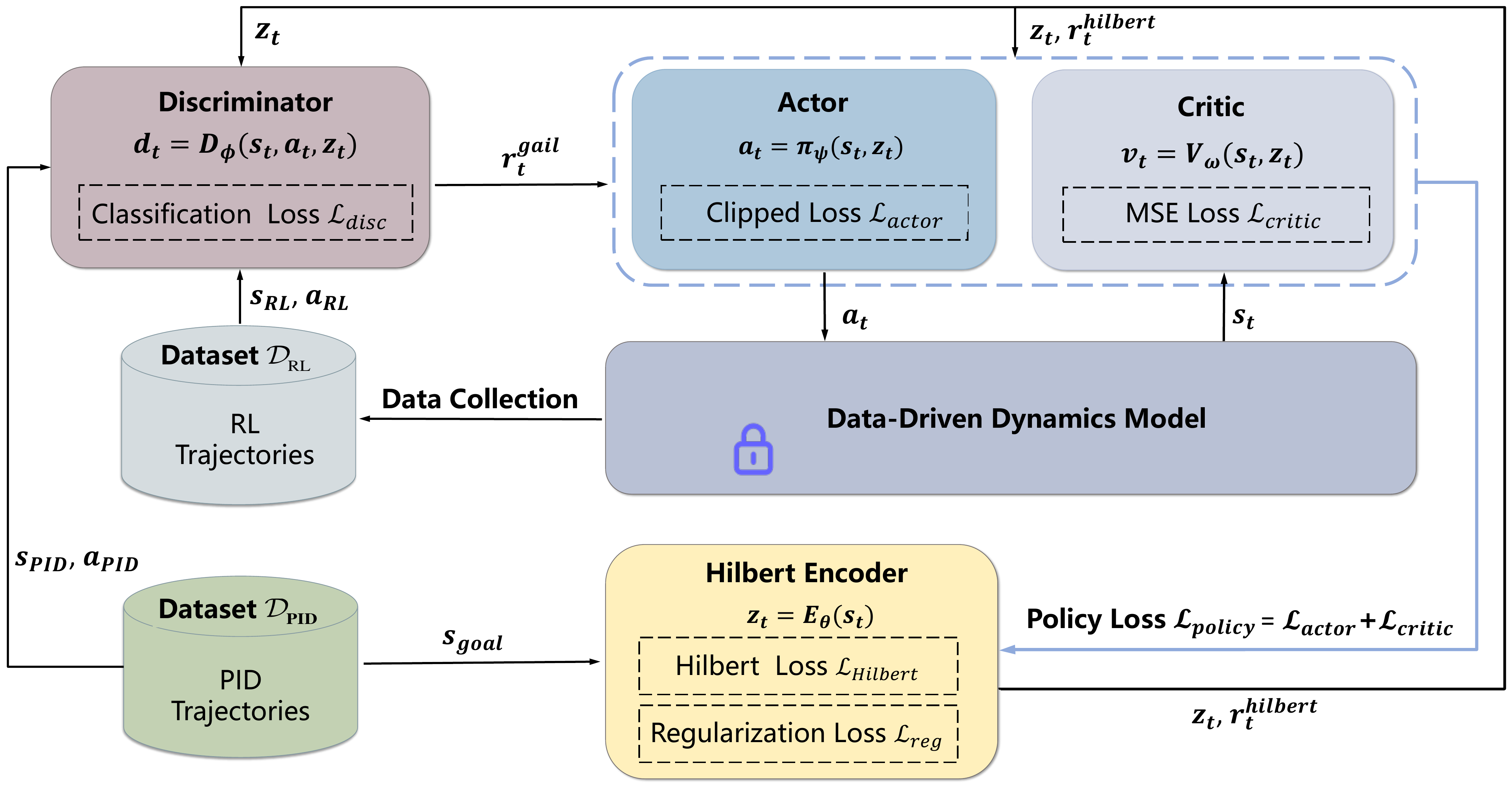}
    \caption{\textbf{Overview of the model training process for the generative RL framework.}
    }
    \label{fig:model_training}
\end{figure*}
Afterward, the actor's objective is defined by the PPO clipped surrogate loss \cite{schulman2017proximal}:
\begin{align}
\mathcal{L}_{\mathrm{actor}}(\psi) &= \mathbb{E} \Biggl[ \min \Biggl( \frac{\pi_{\psi}(a_t|s_t, z_t)}{\pi_{\psi, \mathrm{old}}(a_t|s_t, z_t)}A_t, \nonumber \\
&\quad \mathrm{clip}\left(\frac{\pi_{\psi}(a_t|s_t, z_t)}{\pi_{\psi, \mathrm{old}}(a_t|s_t, z_t)}, 1-\epsilon, 1+\epsilon\right)A_t \Biggr) \Biggr],
\end{align}
where $\epsilon$ is the clipping hyperparameter set to $0.2$. It ensures stability by constraining the policy update size at each step, preventing the new policy $\pi_{\psi}$ from deviating too drastically from the old one $\pi_{\psi, \mathrm{old}}$.
Simultaneously, 
The critic's loss function is defined as:
\begin{equation}
\mathcal{L}_{\rm critic}(\omega) = \mathbb{E} \left[ (V_{\omega}(s_t, z_t) - R_t)^2 \right].
\end{equation}

Finally, we are ready to pre-train the foundation policy, which effectively balances between control stability and efficiency.

\begin{figure*}[ht]
\begin{centering}
\subfigure[]{
    \label{fig:train_loss_encoder}
    \includegraphics[width=0.33\textwidth]{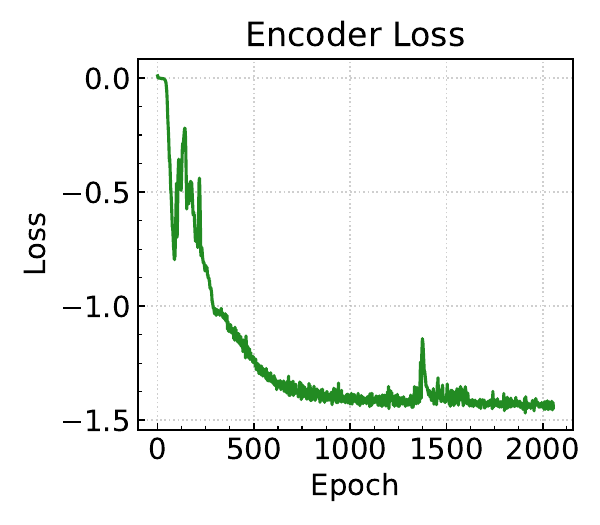}
}
\hspace{-4mm}
\subfigure[]{
    \label{fig:train_loss_discriminator}
    \includegraphics[width=0.33\textwidth]{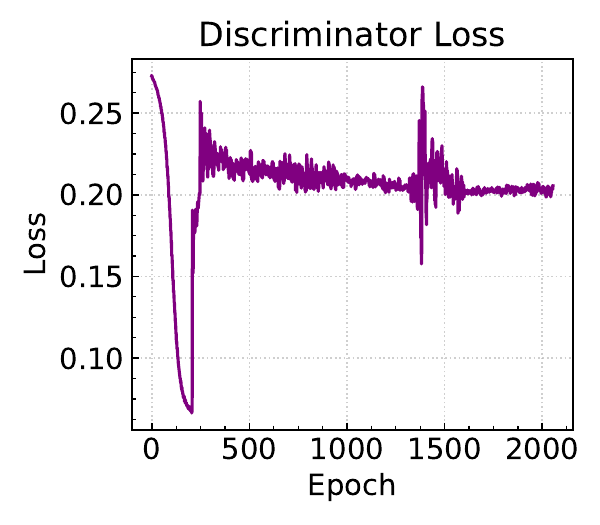}
}
\hspace{-4mm}
\subfigure[]{
    \label{fig:train_score_discriminator}
    \includegraphics[width=0.33\textwidth]{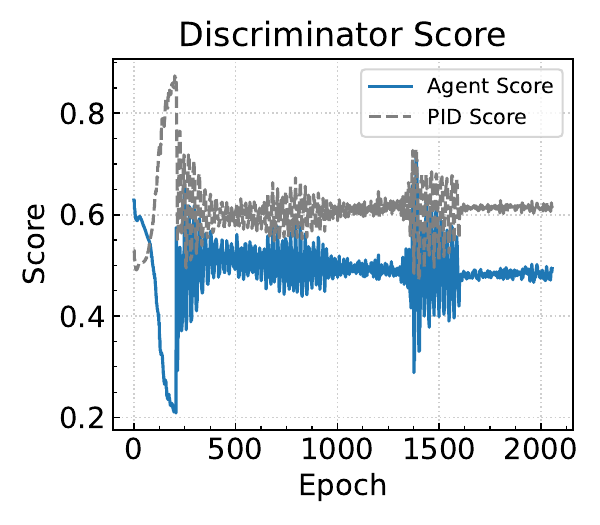}
}
\par\end{centering}
\caption{\textbf{Key performance metrics during the training process.}  This figure illustrates the evolution of critical metrics over 2000 training epochs. (a) Encoder loss, (b) Discriminator loss, and (c) Discriminator scores for both the agent policy and the PID data. }
\label{fig:train_loss}
\end{figure*}
\subsection{Model Training}
\label{Model Training}
\subsubsection{Neural Network Structure}

The core components of our framework, the actor $\pi_{\psi}$, critic $V_{\omega}$, Hilbert encoder $E_{\theta}$, and adversarial discriminator $D_{\phi}$, are all implemented as Multi-Layer Perceptrons (MLPs) featuring ReLU activation functions and Dropout layers with a rate of $0.1$. The detailed architectures for these networks are visually presented in Figure \ref{fig:model_network}. The actor, critic, and discriminator networks utilize initial, separate MLP pathways for their distinct input components. 
These processed representations are then concatenated before being fed into subsequent shared hidden layers for integrated feature learning and decision-making.
The hidden layer width for the actor, critic, and encoder is uniformly set to $128$, as is the width for the discriminator, while the encoder outputs a latent variable $z$, with a dimensionality of $16$. The Adam optimizer \cite{kingma2015adam} is utilized for training, coupled with an exponential learning rate decay schedule to facilitate convergence in the later stages. 
To ensure training stability of PPO, the policy network operates with a fixed logarithmic standard deviation of $-0.5$. 
A comprehensive summary of the model hyperparameters is provided in Table~\ref{tab:model_hyperparameters_full}.

\subsubsection{Training Process}
The training process itself is conducted within a pre-validated, high-fidelity dynamics model from our prior work \cite{wu2025high}. The model has been successfully validated in supporting various plasma control experiments using RL on the actual HL-3 tokamak device. To mitigate the effects of the dynamics model's inherent reality gap and potential error accumulation over long horizons, each interaction trajectory between the generative RL agent and the dynamics model in this work is limited to $32$ time steps. 

An overview of the training process is provided in Figure~\ref{fig:model_training}. The framework is trained in a unified, end-to-end manner. Each training epoch starts with the agent interacting with the dynamics model to collect an experience buffer of $4096$ trajectories, each spanning $32$ steps. To enhance training efficiency, we employ a Distributed Data Parallel (DDP) \cite{li2020pytorch} scheme, conducting synchronous training across eight NVIDIA 4090 GPUs. To initialize each trajectory, a discharge is randomly sampled from the offline dataset $\mathcal{D}_{\rm PID}$, and its plasma parameters at a random time step are used as the initial state. To enhance policy generalization, the corresponding goal state is then selected from a subsequent time step within the same trajectory with an probability of $80\%$ or from a different, randomly chosen trajectory with a probability of $20\%$. The goal state is processed by the Hilbert encoder $E_{\theta}$ to yield a latent variable, which is periodically resampled every $5$ to $10$ steps throughout the trajectory interaction. 
The collected on-policy interaction data $\mathcal{D}_{\rm RL}$ is then paired with samples from the offline dataset $\mathcal{D}_{\rm PID}$ and partitioned into mini-batches of size $16,384$. The losses for the actor, critic, and Hilbert encoder are integrated into a unified objective function for joint optimization. And the discriminator is updated independently using a least-squares loss with a gradient penalty (LSGAN) \cite{mao2017least, gulrajani2017improved} to enhance training stability.

\subsubsection{Convergence of Training Loss}
The key loss functions and performance metrics throughout the training process are visualized in Figure~\ref{fig:train_loss}. The favorable convergence of these metrics systematically validates the effectiveness of our proposed framework.
Figure~\ref{fig:train_loss_encoder} illustrates the Hilbert encoder loss $\mathcal{L}_{\rm Enc}(\theta)$. 
The curve shows that the encoder loss decreases rapidly during the initial training phase and eventually converges to a stable, low value. 
Figure~\ref{fig:train_loss_discriminator} presents the discriminator loss $\mathcal{L}_\text{Disc}(\phi)$, while Figure~\ref{fig:train_score_discriminator} displays the scores assigned by the Discriminator to the agent (agent score) and PID (PID score) behaviors. 
In the early stages of training (approximately the first $200$ epochs), the agent's policy is nearly random, allowing the discriminator to easily differentiate its behavior from the patterns within the PID data, causing the loss to decrease rapidly. Meanwhile, a sharp increase in the PID score and a brief dip in the agent score reflect its quickly established classification capability. However, as the agent's policy improves, the discriminator's task becomes more challenging, causing the loss to rise and eventually stabilize at a dynamic equilibrium point of approximately $0.2$. 
Ultimately, the PID score maintains around $0.6$ and the agent score fluctuates near $0.5$. This result demonstrates the success of the adversarial training, signifying that the RL policy has effectively learned the key control patterns and stable characteristics inherent in the offline PID data.

\section{Results and Analysis}
\label{sec:results}
\subsection{Zero-Shot Trajectory Tracking Performance}
\begin{figure*}[ht]
\begin{centering}
\subfigure[]{
    \label{fig:5020_1200_1700}
    \includegraphics[width=0.33\textwidth]{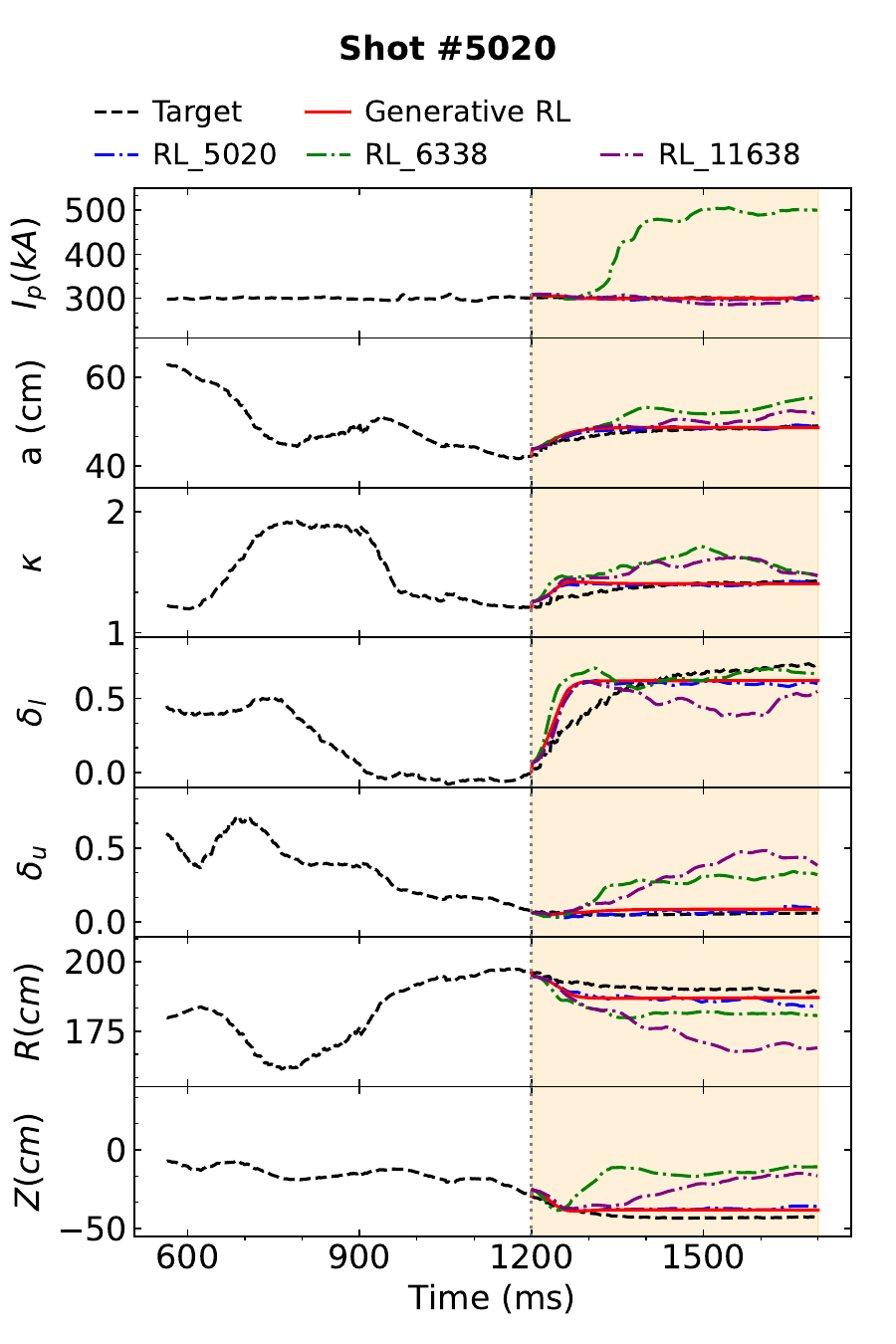}
}
\hspace{-4mm}
\subfigure[]{
    \label{fig:6338_700_1200}
    \includegraphics[width=0.33\textwidth]{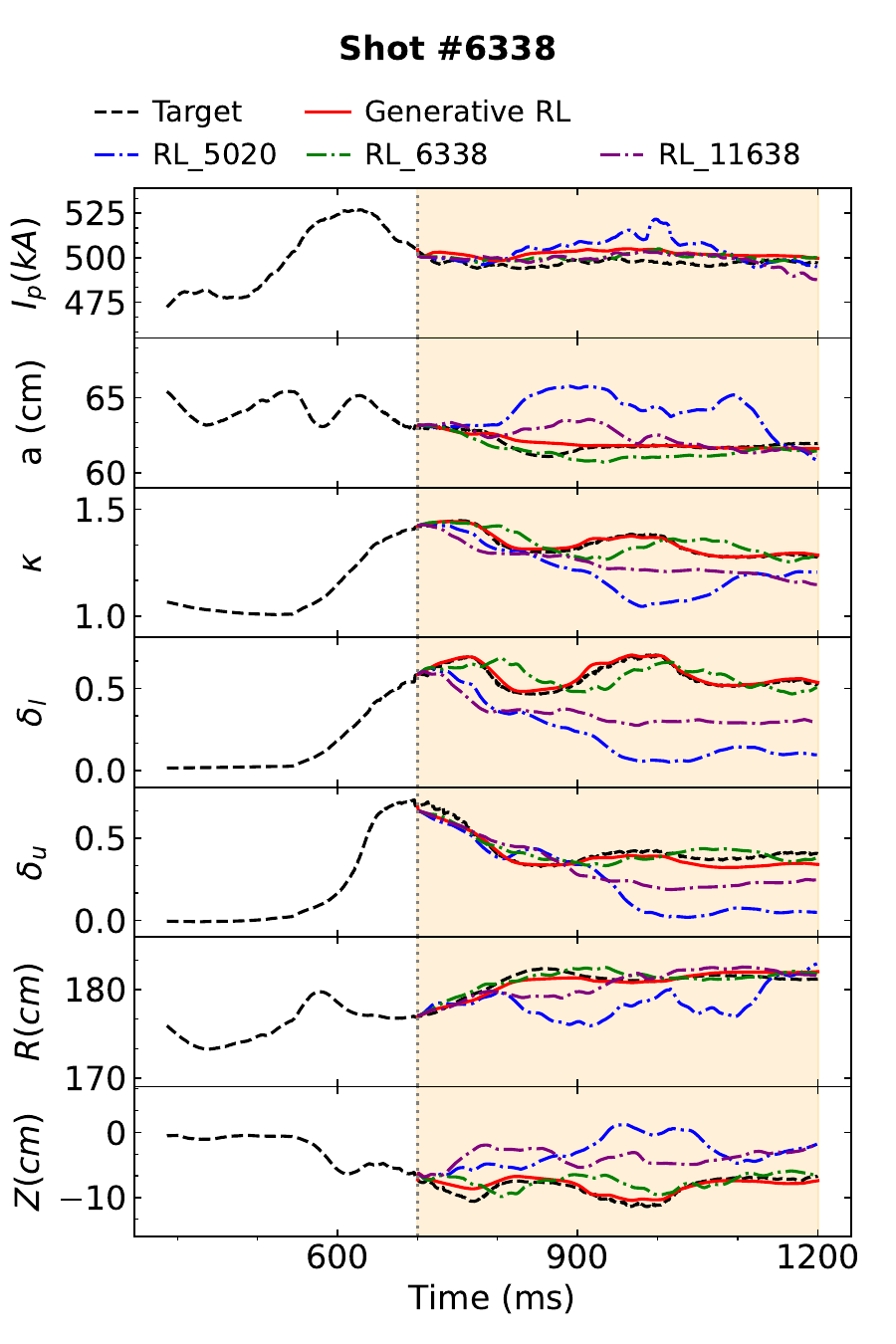}
}
\hspace{-4mm}
\subfigure[]{
    \label{fig:11638_700_1200}
    \includegraphics[width=0.33\textwidth]{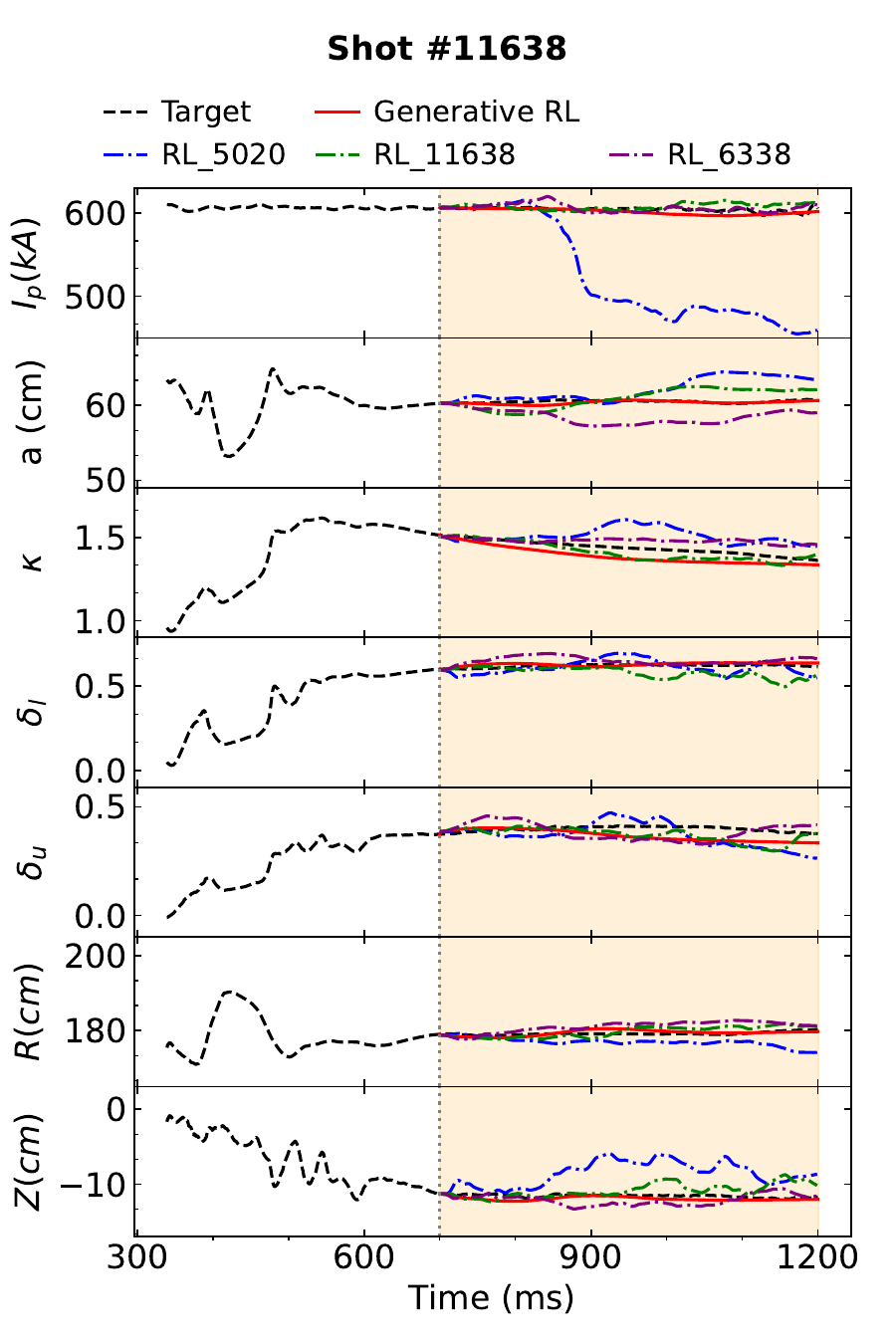}
}
\par\end{centering}
\caption{\textbf{Comparison of zero-shot control performance between the proposed generative RL controller and specialized RL controllers across diverse plasma scenarios. } This figure illustrates the control efficacy of the RL controller in three distinct scenarios without any fine-tuning: (a) shot \#5020, with a plasma current of $300$ kA and a control phase from $1200$ ms to $1700$ ms; (b) shot \#6338, with a plasma current of $500$ kA and a control window from $700$ ms to $1200$ ms; and (c) shot \#11638, with a plasma current of $600$ kA and a control period from $700$ ms to $1200$ ms. In each scenario, the performance of the single, pre-trained generative RL policy (Generative RL) is compared against three specialized policies (RL\_5020, RL\_6338, RL\_11638). Each specialized policy is trained exclusively on the target trajectory of its corresponding shot. 
}
\label{fig:5020_6338_11638}
\end{figure*}

\begin{table*}[htbp]
\centering
\caption{The MAE for all models. The best results are given in bold, while the second best results are underlined.}
\label{tab:control_error}
\resizebox{\textwidth}{!}{%
\begin{tabular}{llcccc}
\toprule
\textbf{Test Shot} & \textbf{Variable} & \textbf{Generative RL} & \textbf{Specialized RL\_5020} & \textbf{Specialized RL\_6338} & \textbf{Specialized RL\_11638} \\
\midrule
\multirow{7}{*}{\textbf{Shot \#5020}} & $I_p$ (kA) & \textbf{1.70} & \underline{1.88} & 133.18 & 7.12 \\
 & $a$ (cm) & \underline{0.70} & \textbf{0.06} & 3.66 & 1.77 \\
 & $k$ & \textbf{0.035} & \underline{0.036} & 0.173 & 0.137 \\
 & $\delta_l$ & \underline{0.091} & 0.092 & \textbf{0.086} & 0.177 \\
 & $\delta_u$ & \underline{0.027} & \textbf{0.022} & 0.188 & 0.230 \\
 & $R$ (cm) & \textbf{3.21} & \underline{3.27} & 8.16 & 14.22 \\
 & $Z$ (cm) & \textbf{3.98} & \underline{4.38} & 24.05 & 15.83 \\
\midrule
\multirow{7}{*}{\textbf{Shot \#6338}} & $I_p$ (kA) & 5.10 & 7.40 & \textbf{2.40} & \underline{3.74} \\
 & $a$ (cm) & \textbf{0.20} & 2.18 & \underline{0.44} & 0.69 \\
 & $k$ & \textbf{0.007} & 0.122 & \underline{0.041} & 0.084 \\
 & $\delta_l$ & \textbf{0.014} & 0.340 & \underline{0.074} & 0.224 \\
 & $\delta_u$ & \textbf{0.032} & 0.201 & \underline{0.041} & 0.125 \\
 & $R$ (cm) & \textbf{0.46} & 2.71 & \underline{0.61} & 1.04 \\
 & $Z$ (cm) & \textbf{0.66} & 5.42 & \underline{1.30} & 4.66 \\
\midrule
\multirow{7}{*}{\textbf{Shot \#11638}} & $I_p$ (kA) & \textbf{3.10} & 80.10 & \underline{3.63} & 3.90 \\
 & $a$ (cm) & \textbf{0.20} & 1.64 & 2.03 & \underline{0.89} \\
 & $k$ & \underline{0.049} & 0.076 & \textbf{0.043} & \textbf{0.043} \\
 & $\delta_l$ & \textbf{0.012} & \underline{0.031} & \underline{0.031} & 0.038 \\
 & $\delta_u$ & \textbf{0.036} & 0.047 & \underline{0.045} & \textbf{0.036} \\
 & $R$ (cm) & \textbf{0.61} & 2.19 & 2.02 & \underline{1.62} \\
 & $Z$ (cm) & \textbf{0.37} & 2.97 & 1.01 & \underline{0.94} \\
\bottomrule
\end{tabular}
}
\end{table*}

\begin{figure}[ht]
    \centering
    \includegraphics[width=0.8\columnwidth]{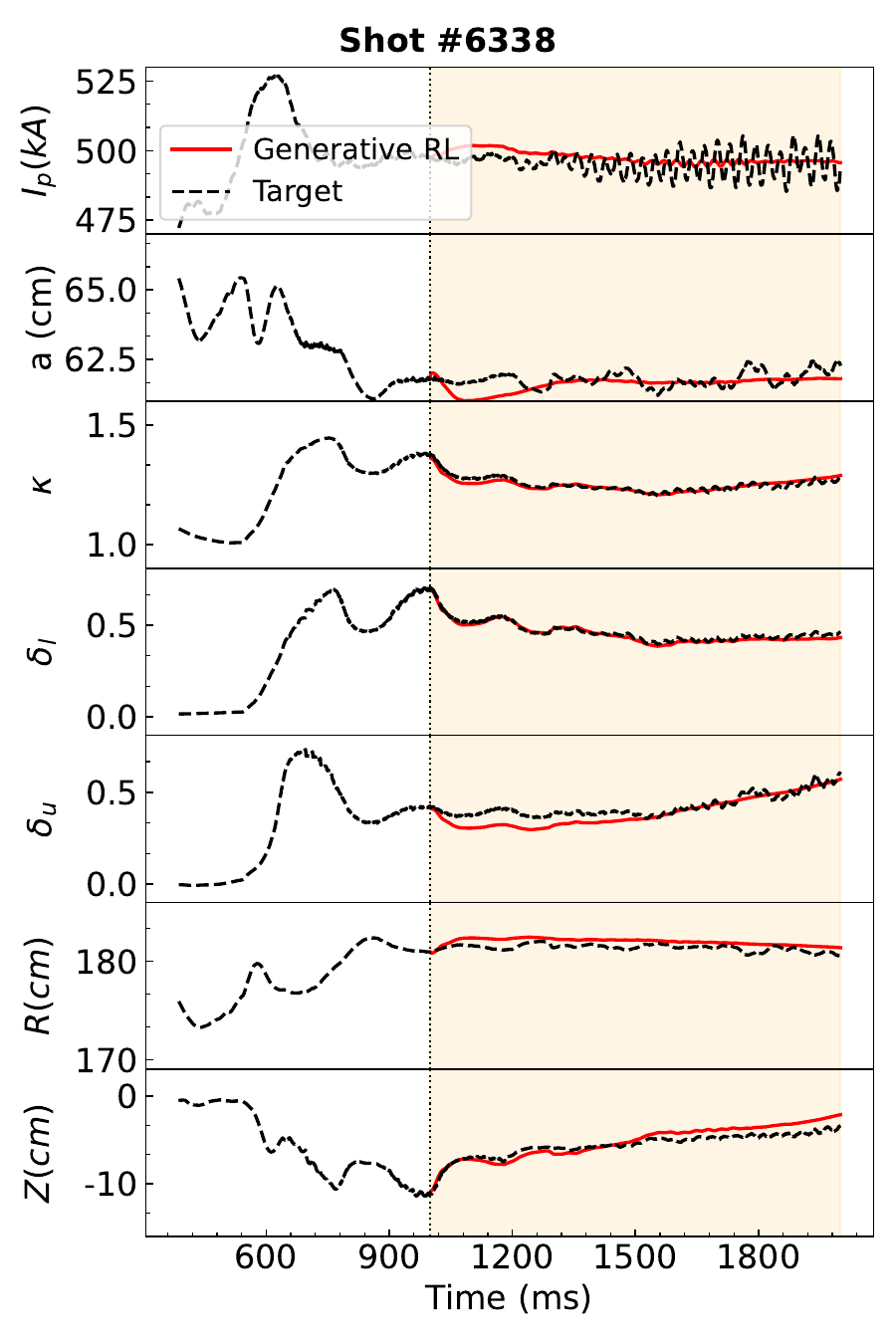}
    \caption{\textbf{Zero-shot control results of the proposed RL controller for shot \#6338.} This figure illustrates the control efficacy of the RL controller for shot \#6338, characterized by a plasma current of $500$ kA with the control phase from $1000$ ms to $2000$ ms. 
    }
    \label{fig:6338_1000_2000}
\end{figure}

\begin{figure*}[ht]
    \centering
    \includegraphics[width=0.9\textwidth]{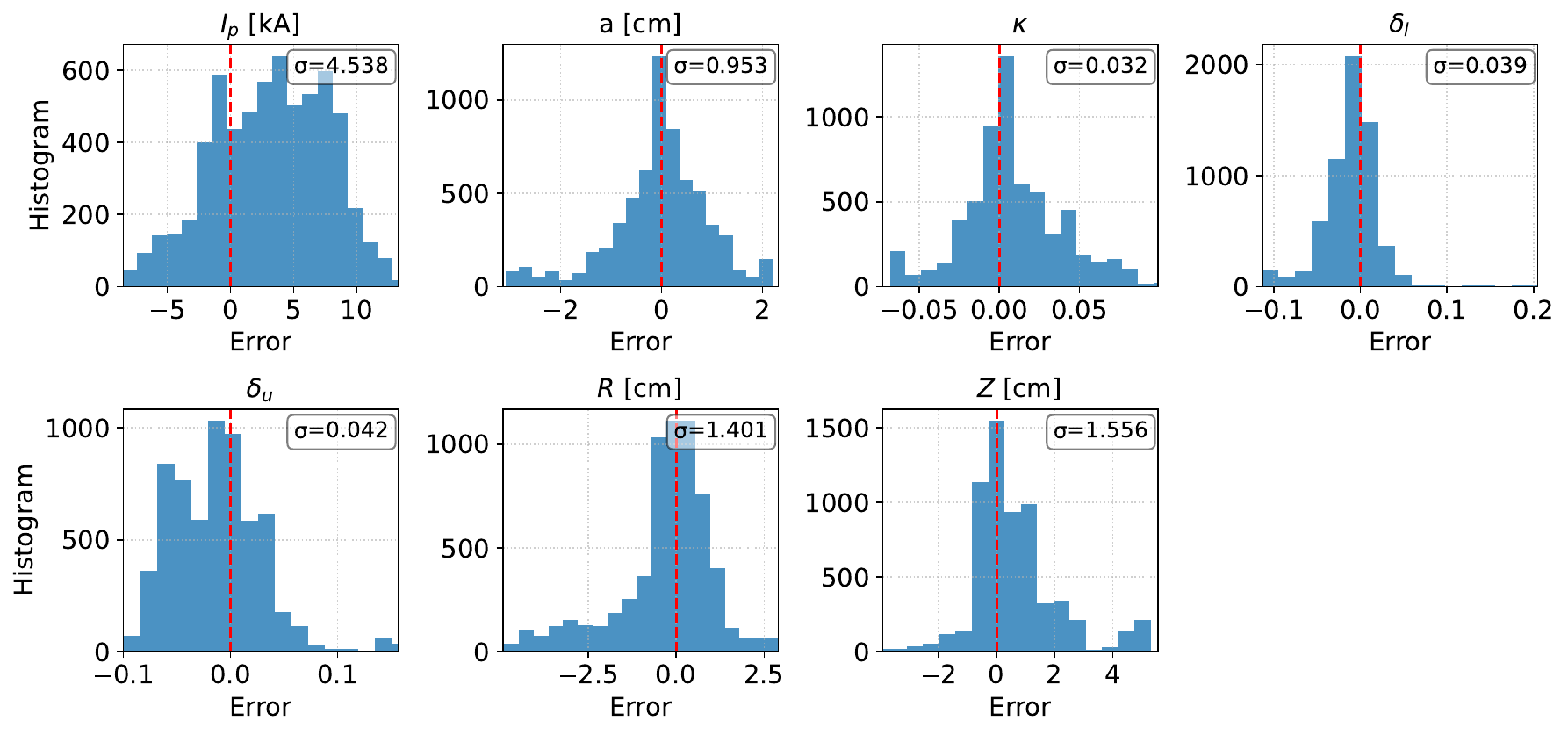}
    \caption{\textbf{Statistical distribution of control errors for plasma current and shape parameters across 10 randomized discharges.} }
    \label{fig:error_histograms}
\end{figure*}

To comprehensively evaluate the control performance of our proposed framework, we conduct a series of closed-loop tracking control tests on a high-fidelity data-driven simulator \cite{wu2025high}. As the dynamics model \cite{wu2025high} has previously validated in online experiments for control durations of approximately $400$ ms, the control duration for these tests is conservatively set to $500$ ms to ensure reliability. 
As illustrated in Figure~\ref{fig:framework}, the pre-trained policy network and Hilbert encoder are loaded to execute closed-loop control tasks in the simulation. 

To ensure the physical viability of the test objectives, we randomly select trajectory segments from historical PID-controlled discharges to serve as control targets. The specific testing procedure is as follows: to provide a proactive control objective, at each time step $t$, we select a future state at a future time $t+\Delta t$ from the reference trajectory as the target state, denoted as $s_{\text{goal},t+\Delta t}$. Afterward, a Hilbert encoder generates the corresponding target latent vector $z_{t}$, where the lookahead window $\Delta t$ is set to $10$ in the experiments. Subsequently, the policy network receives the current plasma state $s_t$ and the target latent vector $z_t$ as input to produce the voltage actions $a_t$ for the $17$ control coils. These actions are then applied to the dynamics model to evolve the system to the next state $s_{t+1}$ , and this cycle is repeated until the task concludes. Figure~\ref{fig:5020_6338_11638} provides tracking control results under three typical plasma current conditions of $300$ kA (shot \#5020), $500$ kA (shot \#6338), and $600$ kA (shot \#11638). 
In all test cases, the agent accurately controls seven parameters simultaneously, including plasma current $I_p$, major radius $R$, minor radius $a$, vertical position $Z$, elongation $\kappa$, and upper and lower triangularity $\delta_u$, $\delta_l$. 
It can be observed that, in shots \#5020 and \#6338, when the target minor radius, elongation, and triangularities undergo near-step changes concurrently, the generative RL agent responds rapidly and converges stably to the target values, showcasing its ability to handle multivariable, strongly-coupled control problems. Shot \#11638 further demonstrates the agent's proficiency in steady-state regulation, precisely maintaining the plasma configuration along the near-constant reference targets. The results demonstrate the zero-shot control capability of the pre-trained policy for reference targets across different current platforms.
To comprehensively benchmark our proposed generative model (Generative RL), we train three additional specialized RL policies: RL\_5020, RL\_6338, and RL\_11638. Following the RL control methodology detailed in our prior work \cite{wu2025high}, each policy is trained exclusively on the target trajectory of its corresponding shot. The comparison in Figure~\ref{fig:5020_6338_11638} clearly shows that while each specialized policy excels on its native task (e.g., the RL\_5020 policy on shot \#5020), its performance degrades sharply when applied to unseen target trajectories. For instance, the policies trained for shots \#6338 and \#11638 exhibit significant tracking errors on multiple parameters when tested on shot \#5020. In contrast, our single generative policy demonstrates stable, high-precision tracking across all three scenarios. For a more direct quantitative comparison, Table~\ref{tab:control_error} summarizes the Mean Absolute Error (MAE) for all control parameters.
 
To further investigate the model's performance limits and its capability for mid-discharge control takeover, an additional long-duration control test is conducted for Shot \#6338, starting at $1000$ ms and concluding at $2000$ ms for a total duration of $1000$ ms, as shown in Figure~\ref{fig:6338_1000_2000}. It can be observed that the agent maintains stable tracking for the entire $1000$ ms duration and continues to closely follow the target trajectory even when it exhibits more complex oscillations in the latter half ($>1500$ ms). This experiment further validates the feasibility for handling unpredictable targets.

Figure~\ref{fig:error_histograms} quantitatively evaluates the overall performance of the RL controller. The analysis compiles tracking errors from $10$ randomized discharges, encompassing both divertor and limiter configurations during the flat-top phase at three distinct plasma current levels: $300$ kA, $500$ kA, and $600$ kA. The results demonstrate high precision in maintaining the plasma's central position. The standard deviations of the error for the horizontal ($R$) and vertical ($Z$) positions of the magnetic axis are $1.401$ cm and $1.556$ cm, respectively. Notably, the error distribution for the horizontal position is tightly centered around zero, indicating an absence of systematic offset in this control dimension. The strategy also exhibits excellent performance in the more intricate task of plasma shape control. The error standard deviations for the minor radius ($a$) and elongation ($\kappa$) are as low as $0.953$ cm and $0.032$, respectively. Furthermore, stable control of the lower ($\delta_{l}$) and upper ($\delta_{u}$) triangularities is achieved, with standard deviations of $0.039$ and $0.042$. Regarding the key global parameter, the plasma current ($I_p$), the error is also effectively controlled with a standard deviation of $4.538$ kA. 
These test results reaffirm that the pre-training framework yields a general-purpose foundation policy capable of multivariable, zero-shot tracking control for arbitrary, physically plausible trajectories by setting goals in a structured latent space.
\begin{figure}[h!]
    \centering
    \includegraphics[width=\columnwidth]{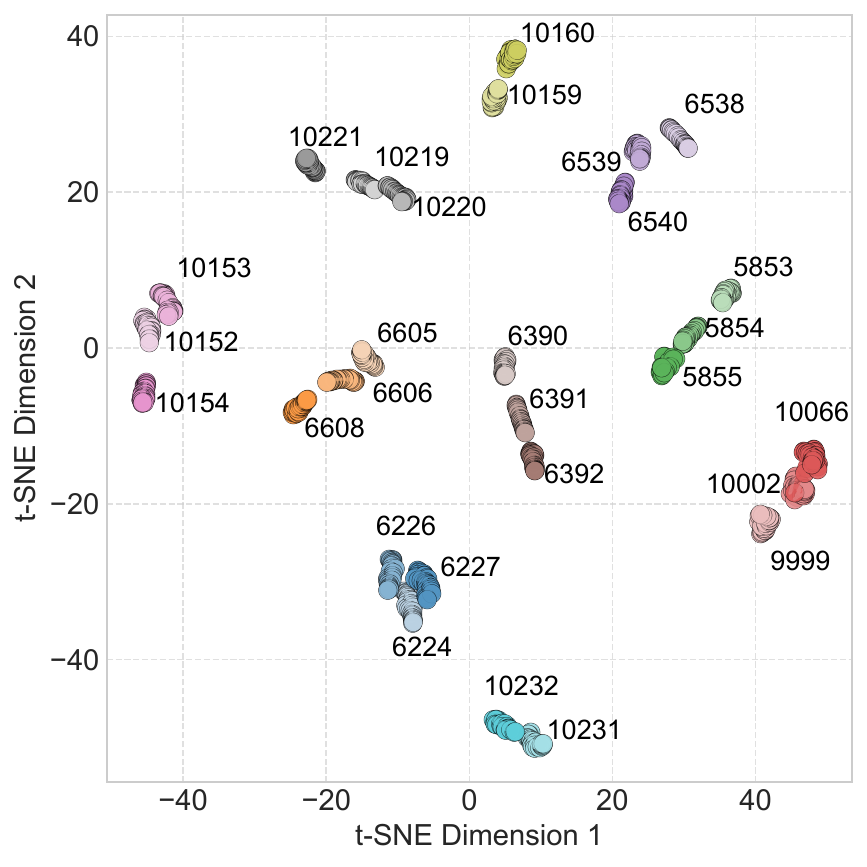}
    \caption{\textbf{t-SNE visualization of the learned Hilbert latent space.}
    }
    \label{fig:t-sne}
\end{figure}

\subsection{Analysis of Learned Hilbert Latent Space}
To evaluate the representational quality of the trained Hilbert encoder, we visualize the generated latent space using t-distributed stochastic neighbor embedding (t-SNE) \cite{maaten2008visualizing}, a dimensionality reduction technique well-suited for high-dimensional data.
The experimental dataset comprises $10$ distinct discharge templates, each with significant configurational differences and providing $2$ to $3$ repeated discharges with similar plasma currents and configurations. From each discharge trajectory, we sample a sequence of $30$ consecutive plasma states $\left\{s_t\right\}_{t=1}^{30}$, which are then mapped by the Hilbert encoder $E_\theta$ into a corresponding sequence of latent vectors  $\left\{z_t\right\}_{t=1}^{30}$. The results are presented in Figure~\ref{fig:t-sne}. 
The t-SNE visualization reveals that the encoder learns a latent space with a well-defined structure, where the Euclidean distance corresponds to the physical similarity of plasma states. Prominently, three key insights can be observed. 
First, clear inter-cluster separability can be observed, where states from the $10$ different templates form distinct clusters in one-to-one correspondence with their source template. Second, strong intra-cluster cohesion is observed, with latent vectors from repeated discharges of the same template forming dense, tightly aggregated subsets. Finally, the mapping preserves temporal continuity, as the consecutive states from each shot form smooth trajectories, ensuring that smooth changes in the input correspond to smooth transitions in the latent space. 
This learned metric structure enables the policy to interpolate or infer a reasonable control output for a new target by evaluating the relative position and distance of its latent vector to those of learned vectors, which is a necessary prerequisite for achieving zero-shot generalization.

\section{Conclusion and Future Research}
\label{sec:discussion}
This work proposes and validates a novel pre-training framework for generalized control of tokamak plasma current ($I_p$) and shape ($a$, $\kappa$, $\delta_l$, $\delta_u$, $R$ and $Z$), designed to address the core challenges of limited flexibility in traditional PID control and poor generalization in single-task oriented reinforcement learning. By synergistically combining Generative Adversarial Imitation Learning (GAIL) and Hilbert space representation learning, we have successfully pre-trained a single, general-purpose foundation policy from a large-scale, unlabeled dataset of $1,115$ historical PID discharges. Simulation experiments demonstrate that the single, pre-trained policy can achieve zero-shot trajectory tracking for multiple control parameters across diverse plasma current and shape scenarios without any task-specific fine-tuning. In the future, we will endeavor to perform on-device experiment with complex goals, which are not limited to trajectory tracking.

\section{Acknowledgments}
The authors would like to thank the entire HL-3 team for providing experimental data. This work was supported in part by the National Key Research and Development Program of China under Grant 2024YFE03020001 and the Zhejiang Provincial Natural Science Foundation of China under Grant LR23F010005.

\section*{References}
\bibliographystyle{unsrt}
\bibliography{ref}

\end{document}